\pdfoutput=1
\documentclass[aps,prl,twocolumn,superscriptaddress,longbibliography]{revtex4-1}

\usepackage{graphicx}
\usepackage{dcolumn}
\usepackage{bm}
\usepackage{color}
\usepackage[colorlinks=true,allcolors=blue,breaklinks=true]{hyperref}
\usepackage{amsmath}
\usepackage[normalem]{ulem}

\begin{document}

\title{Controlling crackling dynamics by triggering low intensity avalanches}

\author{Jonathan Bar\'{e}s}
\email{jb@jonathan-bares.eu}
\affiliation{Laboratoire de M\'{e}canique et G\'{e}nie Civil, UMR 5508 CNRS-University Montpellier, 34095 Montpellier, France}

\author{Daniel Bonamy}
\email{daniel.bonamy@cea.fr}
\affiliation{Service de Physique de l’\'{E}tat Condens\'{e}e, CEA, CNRS, Universit\'{e} Paris-Saclay, CEA Saclay 91191 Gif-sur-Yvette Cedex, France}

\date{\today}

\begin{abstract}
We examine the effect of small, spatially localized, excitations applied periodically in different manners, on the crackling dynamics of a brittle crack driven slowly in a heterogeneous solid. When properly adjusted, these excitations are observed to radically modify avalanche statistics and considerably limit the magnitude of the largest events. Surprisingly, this does not require information on the front loading state at the time of excitation; applying it either at a random location or at the most loaded point gives the same results. Subsequently, we unravel how the excitation amplitude, spatial extent and frequency govern the effect. We find that the excitation efficiency is ruled by a single reduced parameter, namely the injected power per unit front length; the suppression of extreme avalanches is maximum at a well-defined optimal value of this control parameter. This analysis opens a new way to control largest events in crackling dynamics. Beyond fracture problems, it may be relevant for crackling systems described by models of the same universality class, such as the wetting of heterogeneous substrates or magnetic walls in amorphous magnets.
\end{abstract}

\maketitle


Many systems crackle \cite{sethna2001_nature}: When submitted to slow continuous driving, they respond via random impulse-like events, referred to as avalanches, spanning a variety of scales. These systems encompass a large diversity of phenomena as e.g. fracture \cite{maloy2006_prl,bares2014_prl,laurson2013_natcom, bares2018_natcom}, damage \cite{petri1994_prl,ribeiro2015_prl}, imbibition \cite{planet2009_prl,clotet2016_pre}, or plasticity \cite{liu2016_prl,bares2017_pre,papanikolaou2012_nat}. These crackling dynamics come with rare extreme events that can have devastating effects as for the case of earthquakes or avalanches. Since the occurrence of these large events are, so far, impossible to predict, it is paramount to reduce their intensity.

Concerning seismicity, it is well documented \cite{johnson2005_nat,johnson2008_nat,brodsky2014_ar} that gentle local excitations can induce earthquakes even far from the excitation point. Regarding snow hazard in mountains, various devices have been designed to trigger future avalanches in advance by perturbing locally the snowpack with energy impulses. Taking these ideas a step further, one may ask to what extent avalanche statistics in crackling systems is changed by periodically injecting small amounts of energy at the right place?

In this letter, we examine this question in the problem of an interfacial crack driven in a heterogeneous solid, which is an archetype of crackling system \cite{maloy2006_prl,bonamy2008_prl,laurson2010_pre,bares2013_prl,janicevic2016_prl}. Different ways are implemented to inject periodically and in a controlled manner small amounts of energy for excitation. In some cases, clear effects are observed: the avalanche size statistics and inter-event time distribution are radically modified with a severe decrease of the largest events occurrence. In return, numerous avalanches of smaller sizes are triggered by the excitations. Surprisingly, the effect is as large when the excitation location is randomly selected as when it is judiciously chosen at the most loaded point. Conversely, the excitation efficiency is ruled by a single parameter intimately mingling the excitation amplitude, depth and frequency. The suppression of the extreme avalanches is maximum at a well-defined optimal value of this control parameter. These observations opens a new way to control largest events in crackling dynamics.


\begin{figure}
\centering
\includegraphics[width=0.72\linewidth]{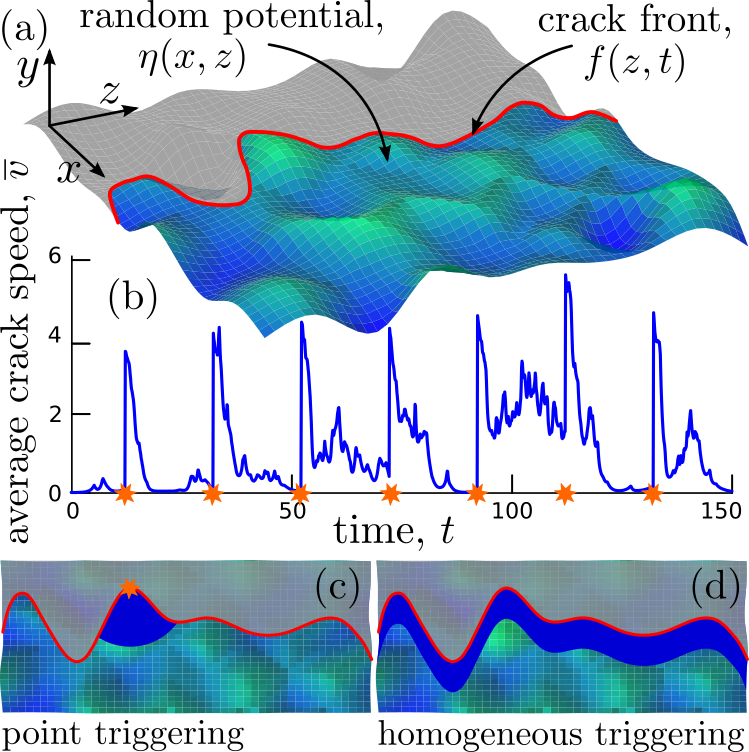}
\caption{(color online) (a) Schematic view of an elastic manifold (red line) $f(z,t)$ driven along the $x$-axis direction in a random potential. The grey part shows the broken area while the random potential is represented with relief and color code (from blue/low to green/high). (b) Typical evolution of mean crack speed, $\overline{v}(t)$, for a simulation with $c=10^{-4}$, $k=10^{-2}$, $N=1024$, $\mathcal{P}=2$, $\mathcal{T}=20$ and $\mathcal{A}=200$. External excitations are indicated by orange stars. (c) Schematic view of the modification imposed to the random map $\eta(z,x)$ when a spatially localized excitation is prescribed (methods 1, 2 and 3). Orange star indicates the triggering point $(z_e,x_e)$. (d) Schematic view of the modification imposed to $\eta(z,x)$ when a homogeneous excitation is prescribed (method 4). In both panels (c) and (d), dark blue areas correspond to the zones where $\eta(z,x)$ is set to $\mathcal{P}$.}
\label{fig_1}
\end{figure}

\noindent\emph{Numerical method --} Crack growth in brittle heterogeneous solids can be mapped to a long-range elastic spring driven in a random potential \cite{gao1989_jam,schmittbuhl1995_prl,ramanathan1997_prl} [Fig. \ref{fig_1}(a)], and the crackling dynamics sometimes observed is then attributed to the self-adjustment of the driving force around its depinning value \cite{bonamy2008_prl,laurson2010_pre,laurson2013_natcom}. The derivation of the equation of motion for the crack line, $f(z,t)$, were detailed elsewhere \cite{bares2013_prl,bares2014_ftp,bares2019_pre}; it writes:
\begin{equation}
\begin{split}
	\frac{\partial f}{\partial t} = ct-k\overline{f} + J(z,\{f\}) + \eta(z,x=f(z,t),t)\\
	\mathrm{with}\quad J(z,\{f\})=\dfrac{1}{\pi} \times \int{\dfrac{f(\zeta,t)-f(z,t)}{(\zeta-z)^2}d\zeta}
\end{split}
	\label{eqLine}
\end{equation}
\noindent Here, $\vec{e}_x$ and $\vec{e}_z$ axes are aligned with direction of crack propagation and crack front, respectively. $c$ is the driving rate, the rate at which the loading displacement increases; $k$ is the unloading factor, the rate at which the solid stiffness decreases with crack length. The integral term on the right-hand side represents the long-range elastic interactions and translates the local perturbations of stress intensity factor caused by the front distortions \cite{gao1989_jam}. Finally, the random potential $\eta(z,x,t)$ models toughness fluctuations in the material. Note that, here, this term explicitly depends on time contrary to the quenched disorder $\eta(z,x=f(z,t))$ assumed in the common implementation of this model \cite{schmittbuhl1995_prl,ramanathan1997_prl,bonamy2008_prl,laurson2010_pre,bares2013_prl,bares2014_ftp,janicevic2016_prl,bares2019_pre}. This extra dependence in time is used to excite the system as described below.

At periodically distributed times $t_e=\mathcal{T}, 2\mathcal{T}...$, small disturbances tickle crack propagation. Four methods are implemented : {\em First method} (M1) consists in picking {\em randomly} a point $(z_e,x_e=f(z_e,t_e))$ along the front and considering the unbroken material near this point, within a radius $r_{\mathcal{A}}$; there, $\eta$ is arbitrary set to a constant value $\mathcal{P}$ [Fig. \ref{fig_1}(c)]:           
\begin{equation}
	\eta(r,t>t_e)=\mathcal{P} ~~~~ \forall r<r_{\mathcal{A}},
	\label{eqExcit1}
\end{equation}
\noindent where $r=\sqrt{(z-z_e)^2+(x-x_e)^2}$ with $x>f(z,t_e)$.  In addition to $\mathcal{P}$, the excitation amplitude is set by parameter $\mathcal{A}$ defined such that $\int_0^{r_{\mathcal{A}}}\eta(r,t_e^-)-\eta(r,t_e^+)dr=\mathcal{A}$ where $t_e^-$/$t_e^+$ denotes the time just before/after the excitation. {\em Second method} (M2) consists in placing $(z_e,x_e)$ at the {\em most loaded} point along the elastic line at time $t_e$, that is the position $(z_e,x_e=f(z_e,t_e))$ where $J(z,\{f\})$ in Eq. \ref{eqLine} is maximum. $\eta(r,t>t_e)$ is then modified following Eq. \ref{eqExcit1} as in M1. In {\em third method} (M3) $(z_e,x_e)$ is placed at the {\em least loaded} point of the elastic line. Finally, in {\em fourth method} (m4), $\eta$ Is set to $\mathcal{P}$ all along the front at $t_e$, within a strip of width $x_a$:
\begin{equation}
	\eta(z,x,t>t_e)=\mathcal{P} ~~~~ \forall x \in [f(z,t_e),f(z,t_e)+x_{\mathcal{A}}],
	\label{eqExcit2}
\end{equation}
\noindent where $x_{\mathcal{A}}$ is defined such that $\int_z \int_{f(z,t_e)}^{f(z,t_e)+x_{\mathcal{A}}}\eta(z,x,t_e^-)-\eta(z,x,t_e^+)dx dz=\mathcal{A}$ [Fig. \ref{fig_1}(d)].

The initial map $\eta(x,z)$ is first prescribed as a $1024$ width uncorrelated random map with zero average and unit variance. Then, Eq. \ref{eqLine} is solved using a fourth order Runge-Kutta scheme on a $2$GHz CPU, as in \cite{bares2013_prl,bares2014_ftp,bares2018_ptrsa}. Note that the long-range kernel in the right-hand side of Eq. \ref{eqLine} takes a simpler expression in the $z$-Fourier space: $\hat{J}(z,\{f\})=-|q|\hat{f}$; hence, periodic conditions along $z$ are prescribed and, $J(z,\{f\})$ is computed in the $z$-Fourier space. Driving rate and unloading factor are fixed to $c=10^{-4}$ and $k=10^{-2}$ respectively. This ensures a clear crackling dynamics with giant velocity fluctuations \cite{bares2013_prl,bares2019_pre} in absence of external perturbations. The parameters defining these latter were varied in the following ranges: $\mathcal{P} \in [1,10]$, $\mathcal{T} \in [0.2,600]$ and $\mathcal{A} \in [2,700]$. 

Figure \ref{fig_1}(b) shows a typical time profile of the spatially-averaged crack speed $\overline{v}(t)=\langle \frac{\partial f}{\partial t} \rangle_z$. As classically done for such systems \cite{bares2014_ftp,bares2018_ptrsa}, the depinning avalanches were identified with the excursions of $\overline{v}(t)$ above a prescribed threshold $\overline{v}_{\rm{th}}$. For each avalanche $i$, occurrence time $t_i$ is defined as the first time at which $\overline{v}(t)$ exceeds $\overline{v}_{\rm{th}}$, duration $D_i$ as the time interval when $\overline{v}(t)$ stays over $\overline{v}_{\rm{th}}$, and size $S_i$ as the integral of $\overline{v}(t)-\overline{v}_{\rm{th}}$ over this time interval. The threshold is set to the average global speed: $\overline{v}_{\rm{th}}=c/k=10^{-2}$. To clearly isolate the effect of the excitation, events are separated between those directly at $t_e$, and the others.


\begin{figure}
\centering
\includegraphics[width=0.85\linewidth]{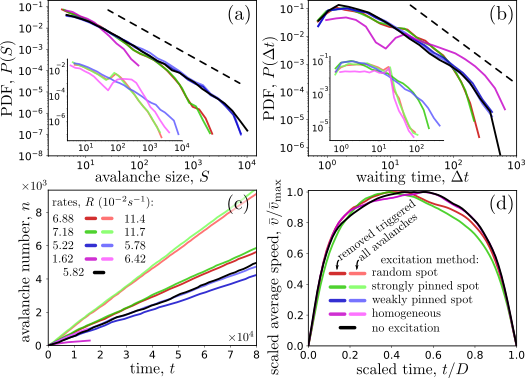}
\caption{(color online) (a) PDF of avalanche size, $P(S)$. Straight dashed line shows a power-law of exponent $\beta=1.3$. (b) PDF of the waiting time between consecutive avalanches of size larger than $4$, $P(\Delta t)$. The straight dashed line shows a power-law with exponent $p=1.75$. In the main panels of both (a) and (b), the avalanches directly triggered by the periodic excitations are withdrawn, while in the insets, PDFs consider all avalanches. (c) Cumulative number of avalanches as a function of time. Slopes give avalanche rates, $R$, provided in inset. (d) Temporal avalanche shape, averaged over all avalanches with duration ranging between $3$ and $5$. In panels (a)-(d), black curve is the reference one in absence of excitation and the colored ones correspond to the four different types of excitations (see inset in panel (d)), pale curves considers all avalanches, and dark ones consider non-triggered ones only. All these curves were obtained with $\mathcal{A}=200$, $\mathcal{T}=20$ and $\mathcal{P}=2$.}
\label{fig_2}
\end{figure}

\noindent\emph{Role of the triggering method --} 
In absence of external excitations, both avalanche size [Fig. \ref{fig_2}(a)] and waiting time [Fig. \ref{fig_2}(b)] exhibit clear scale free statistics, with power-law probability density function (PDF) extending over almost four decades. But, depending on the prescribed method, small intensity disturbances can alter the observed features. As expected, excitations at the least loaded point along the front (M3) has almost no effects, while those generated at the most loaded point (M2) yield drastic effects. More surprisingly, choosing randomly excitation location (M1) yields the same effect as the latter: The excitations brought by M1 and M2 cut the largest avalanches in smaller ones generating a bump in $P(S)$ for $S\sim S_{trig}$ [Inset in Fig. \ref{fig_2}(a)]. This bump corresponds to the avalanches directly triggered by the excitation and disappears when they are removed from the data set. The PDF then turns into a gamma distribution with exponent $\beta=1.3$ and an upper cut-off $S_{\rm{max}}$ greatly reduced [Main panel in Fig. \ref{fig_2}(a)]. Regarding waiting times, the PDF exhibits a maximum at $\Delta t = \mathcal{T}$ when all events are considered [Inset in Fig. \ref{fig_2}(b)]; above $\mathcal{T}$, $P(\Delta t)$ is truncated. Once the triggered events are removed, $P(\Delta t)$ obeys a gamma distribution with $p=1.75$ but with a reduced upper cut-off, $\Delta t_{\rm{max}}$ [Main panel in Fig. \ref{fig_2}(b)]. Note finally that when the excitation stops being spatially localized (M4), $P(S)$ is also significantly modified but the decrease observed in $S_{\rm{max}}$ is much smaller than that for M1 and M2. Regarding waiting time, M4 does not generate significant decrease of $\Delta t_{\rm{max}}$.

The effect of applied disturbances onto avalanche rate is examined in Fig.\ref{fig_2}(c). This rate is computed either by considering all events ($R_{all}$, light colors) or after removing triggered events ($R_{w/ot}$, dark colors). $R_{w/ot}$ increases slightly (of $\sim 20\%$) when excitations are generated with M1 or M2. The rate of triggered avalanches, $R_{all}-R_{w/ot}$, is quite large in both cases. Conversely, $R_{all}-R_{w/ot}$ is quite small when M3 is applied and both are nearly equal to the avalanche rate measured in absence of excitations. Regarding M4, most of the avalanches ($\sim 80\%$) are directly triggered by the excitations.

Finally the temporal avalanche shape was examined. This observable provides an important characterization of crackling signals \cite{laurson2013_natcom,dobrinevski2015_epl,abed2019_pre}. This shape is obtained by ($i$) identifying all avalanches with durations $D_i$ falling into a prescribed interval and then ($ii$) by averaging the shape $\overline{v}^i(t)/\overline{v}^i_\mathrm{max}$ vs. $t/D_i$ over all the collected avalanches. Figure \ref{fig_2}(d) shows the determined shapes for $3\leq D \leq 5$. In absence of disturbances, the shape is nearly parabolic, as already reported \cite{bares2014_prl,bares2019_pre}. It remains nearly unaffected by M3 and M4. Conversely, M1 and M2 have a significant effect and slightly shifted the shape to the left. This is the signature of avalanches where initial acceleration phase is faster than subsequent deceleration. The more pronounced asymmetry observed for M1 and M2 compared to M33 simply reflects the greater effectiveness of the first two methods in triggering avalanches. Triggered avalanches are indeed expected to display an asymmetrical faster acceleration phase since they start from a point weakened by the external disturbance, and terminate out of the perturbed zone. The absence of visible asymmetry for M4 is attributed to the fact that, since disturbances are applied to spatially extended zones, triggered avalanches exhibit both a faster initial acceleration phase and final deceleration phase.


\begin{figure}
\centering
\includegraphics[width=0.85\linewidth]{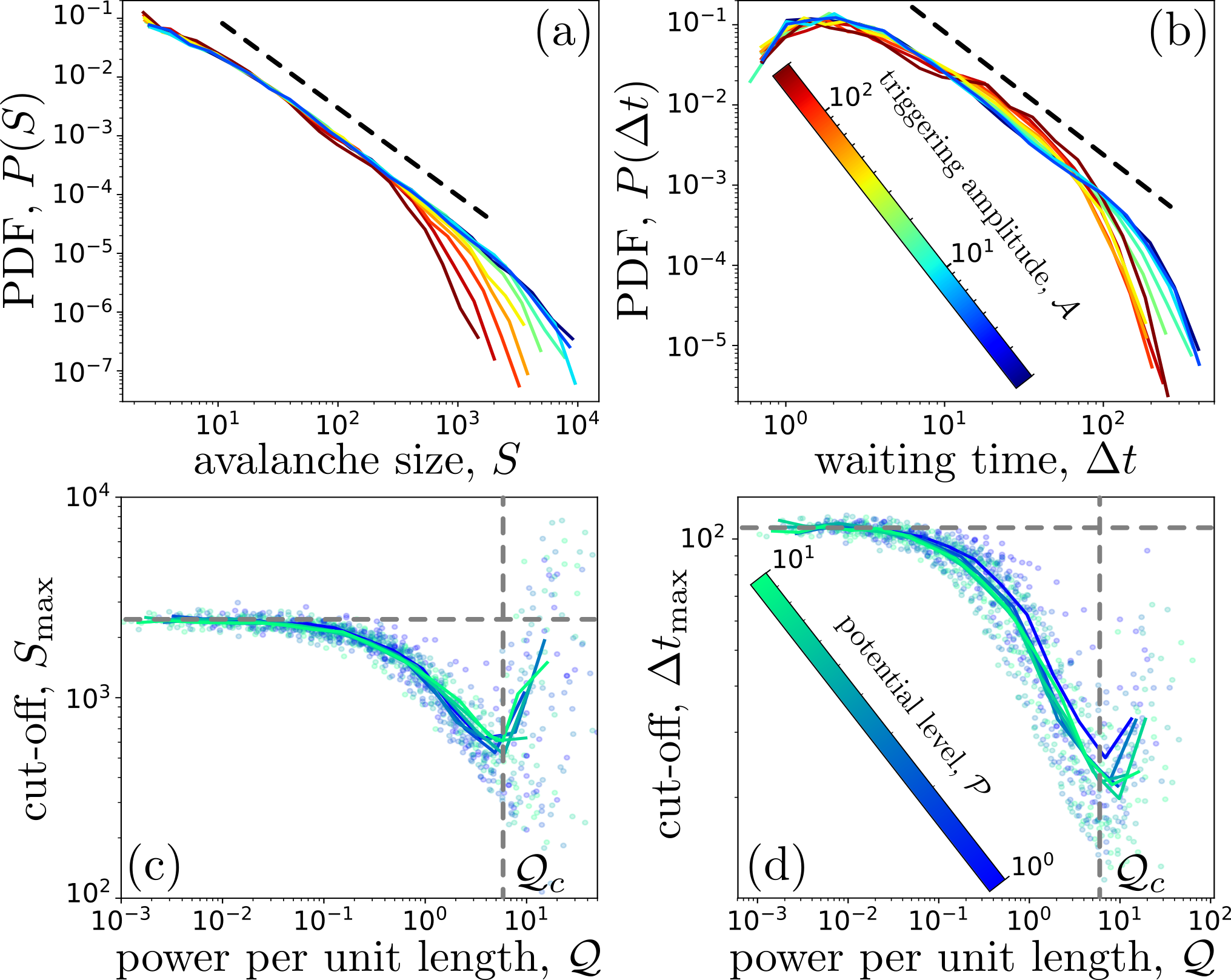}
\caption{(color online) Evolution of the avalanche size PDF, $P(S)$, and waiting time PDF, $P(\Delta t)$, with triggering amplitude, $\mathcal{A}$ are plotted in panels (a) and (b) respectively. In both panels, simulations are done with $\mathcal{T} = 20$ and $\mathcal{P} = 2$ and the PDF decay as a power-law (straight dashed lines) of exponents $\beta=1.3$ (panel (a)) or $p=1.75$ (panel b) up to upper cutoffs $S_\mathrm{max}$ and $\Delta t_\mathrm{max}$ that decrease with increasing $\mathcal{A}$. Variations of $S_\mathrm{max}$ and $\Delta t_\mathrm{max}$ as a function of the power density $\mathcal{Q} = \mathcal{A}/\mathcal{T}/\sqrt{\mathcal{P}}$ are plotted in panels (c) and (d), respectively. In both panels, solid curves show average results at constant $\mathcal{P}$ and vertical dashed lines indicate $\mathcal{Q}_c \sim 6$. Horizontal lines show the cut-off values with no excitation. Statistical analysis were performed on non-triggered avalanches only.}
\label{fig_3}
\end{figure}

\noindent\emph{Role of the triggering parameters --} We now examine more quantitatively the effects of the excitation parameters:  $\mathcal{A}$,  $\mathcal{P}$, and  $\mathcal{T}$. Local excitations applied to a random point (M1) or at the most loaded point (M2) are the most efficient ones and yields same consequences; hence, in the following, only simulations using M1 are examined. Figures \ref{fig_3}(a) and \ref{fig_3}(b) show the evolution of $P(S)$ and $P(\Delta t)$ for increasing excitation amplitude, $\mathcal{A}$, and fixed $\mathcal{T}=20$ and $\mathcal{P}=2$. Both $P(S)$ and $P(\Delta t)$ continue to exhibit a scale-free power-law regime with exponents $\beta = 1.3$ and $p=1.75$ independent of $\mathcal{A}$. Conversely, in both cases, the upper cut-offs, $S_{max}$ and $\Delta t_{max}$ decrease as $\mathcal{A}$ increases, except for the highest values $\mathcal{A}$ where they both increase again with $\mathcal{A}$.

These cutoffs were computed using: $S_{\rm{max}} = \langle S^2 \rangle / \langle S \rangle$ and $\Delta t_{\rm{max}} = \langle \Delta t^2 \rangle / \langle \Delta t \rangle$ \cite{rosso2009_prb,corral2015_csf,bares2019_pre}. {\em A priori}, these two cutoffs depend on $\mathcal{A}$,  $\mathcal{P}$, and  $\mathcal{T}$. But, as shown by the collapses in Figs. \ref{fig_3}(c) and \ref{fig_3}(d), their dependence is fully dictated by a single parameter, $\mathcal{Q} = \mathcal{A} / \mathcal{T} / \sqrt{\mathcal{P}}$. This latter quantifies the injected power per front length unit: $\mathcal{A} / \mathcal{T}$ is the amount of potential injected in the system per unit time and, at fixed $\mathcal{A}$, since $\mathcal{P}$ is proportional to the excited area, $\sqrt{\mathcal{P}}$ scale with the length of the front line in contact with this excited area. For low $\mathcal{Q}$ the upper cut-offs are constant and equal to their value in absence of excitation. Then, from $\mathcal{Q} \approx 10^{-1}$, both $S_{\rm{max}}$ and  $\Delta t_{\rm{max}}$ start decreasing rapidly and reach a minimum at $\mathcal{Q} = \mathcal{Q}_c \approx 6$. There the cutoffs are about five times smaller than what is obtained in absence of excitation. Above $\mathcal{Q}_c$, the cutoffs increase again rapidly. This $\mathcal{Q}_c$ is the optimal injected power to reduce extreme events.

\begin{figure}
\centering
\includegraphics[width=0.65\linewidth]{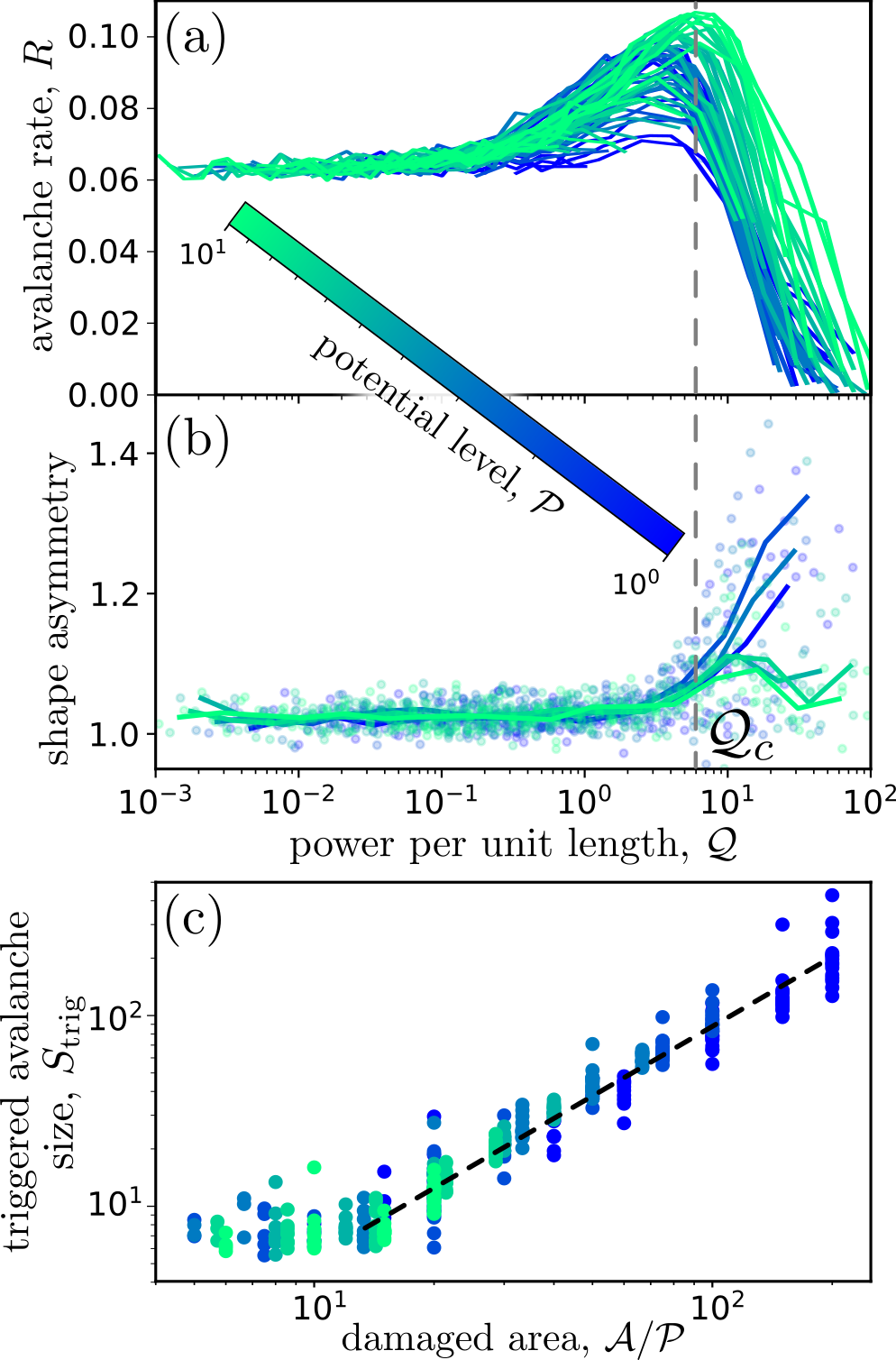}
\caption{(color online) a,b: Number of avalanche per unit time, $R$, and avalanche shape asymmetry as a function of the power density $\mathcal{Q}$. The vertical dashed lined show $\mathcal{Q} \sim 6$. In b solid curves show average results for constant $\mathcal{P}$. c: Averaged triggered avalanche size, $S_{\rm{trig}}$ as a function of the size of the area damaged for triggering, $\mathcal{A}/\mathcal{P}$.}
\label{fig_4}
\end{figure}

The variation of the avalanche rate $R$ with $Q$ is plotted on Fig. \ref{fig_4}(a). $R$ remains roughly constant at low $Q$, and starts increasing rapidly at $\mathcal{Q} \approx 10^{-1}$ to reach a maximum at $\mathcal{Q} = \mathcal{Q}_c$. The range $Q$ where $R(Q)$ shows a bump is the same as the range where $S_{\rm{max}}$ and $\Delta t_{\rm{max}}$ have a dip. The evolution of avalanche shape is finally characterized. As in \cite{laurson2013_natcom,dobrinevski2015_epl}, the asymmetry is defined as the ratio between the integral of $v/\max(v)$ vs. $t/D$ over the left $t/D \in [0,0.5]$ and the right $t/D \in [0.5,1]$ part of the curve (Fig. \ref{fig_2}d). In Fig. \ref{fig_4}b, this asymmetry is plotted for all the simulations as a function of $\mathcal{Q}$. Its value is constant slightly above $1$ at low $Q$ and starts increasing just before $\mathcal{Q}_c$. Finally, in Fig.\ref{fig_4}c, the average size of triggered avalanches, $S_{\rm{trig}}$, is plotted as a function of $\mathcal{A}/\mathcal{P}$, the area of $\eta$ that is modified by the excitation. For damaged areas larger than $10$, the points collapse on a power-law with an exponent slightly (but significantly) larger than 1, close to $1.2$. For $\mathcal{A}/\mathcal{P}$ smaller than $4$, no avalanche is triggered.


\noindent\emph{Concluding discussion -- } In this letter, we examined how small excitations applied periodically during slow crack growth in an heterogeneous solid modifies the crackling dynamics. Localized excitations caused by the weakening of a small area along the crack front can have a considerable effect on avalanche statistics and strongly limit the magnitude of larger events. Not surprisingly, the effect is significant when the weakened zone is chosen at the most loaded point. More surprisingly, the effect is as large when this area is randomly selected along the front. This can be understood by noting that the crack line spends more time on the pinned configurations (the most loaded ones). Therefore, random local excitations mainly tickle the most loaded points.

Applied excitations can {\em a priori} be varied in terms of amplitude ($\mathcal{A}$, quantity of removed potential), depth ($\mathcal{P}$, damaged potential minimum value) or periodicity ($\mathcal{T}$). A second surprising result is that the excitation effect onto the crackling dynamics is fully governed by a single reduced parameter $\mathcal{Q} = \mathcal{A} / \mathcal{T} / \sqrt{\mathcal{P}}$, namely the injected power per unit length of front. When $Q$ is too small, excitations are too weak/rare/spread out to significantly modify the crackling dynamics. When $Q$ is too large, the whole random map ($\eta$) ahead of the front is modified and one hand up with a modified frozen map. But there is a critical value $\mathcal{Q}_c$ where the avalanche statistics is significantly modified and the size of the largest events is greatly reduced. In return, the excitations trigger many additional avalanches, the size of which scale with $\mathcal{A}/\mathcal{P}$. By tuning properly the excitation parameters $\mathcal{A}$, $\mathcal{P}$ and $\mathcal{T}$, it is therefore possible to set $\mathcal{Q} = \mathcal{Q}_c$ so that the largest events in the crackling dynamics are replaced by numerous smaller events of prescribed sizes.

This work opens a new way to control crackling limit inopportune extreme events. Beyond solid failure, our analysis directly extends numerous systems described by the same long-range string model, such as the dynamics of contact lines in wetting problems and the dynamics of domain walls in ferromagnets. As such, it may be directly applied in other fields like nanofluidic or nanomagnetism where crackling and random large scale events are to be limited. More generally, we believe these results can be somehow extended to other crackling systems such as sheared granular matter \cite{bares2017_pre,zadeh2019_pre,zadeh2019_pre_2,hayman2011_pag}, damage \cite{petri1994_prl,ribeiro2015_prl}, neural activity \cite{beggs2003_jn,bellay2015_elf}, human conflicts \cite{richardson1944_jrss,clauset2019_arx} or seismicity \cite{bak2002_prl,davidsen2013_prl} to name a few.


\bibliographystyle{apsrev4-1}
\bibliography{biblio.bib}

\end{document}